\documentclass[twocolumn,english,prl,showpacs,superscriptaddress]{revtex4-1}
\usepackage{epsfig}
\usepackage{graphicx}
\usepackage{amsfonts}
\usepackage[figuresright]{rotating}
\usepackage{amssymb}
\usepackage{amsmath}
\usepackage{amsmath}
\usepackage{graphicx}
\usepackage{color}
\usepackage{hyperref}
\usepackage[final]{pdfpages}

\begin{document}

\title{Disorder effects in correlated topological insulators}
\author{Hsiang-Hsuan Hung}
\affiliation{Department of Physics, The University of Texas at
Austin, Austin, TX, 78712, USA }
\author{Aaron Barr}
\affiliation{Department of Physics, The University of Texas at
Austin, Austin, TX, 78712, USA }
\author{Emil Prodan}
\affiliation{Department of Physics, Yeshiva University, New York, NY 10016, USA}
\author{Gregory A. Fiete}
\affiliation{Department of Physics, The University of Texas at
Austin, Austin, TX, 78712, USA }
\begin{abstract}
Using exact diagonalization and quantum Monte Carlo calculations we investigate the effects of disorder on the phase diagram of both non-interacting and interacting models of two-dimensional topological insulators.  In the fermion sign problem-free interacting models we study, electron-electron interactions are described by an on-site repulsive Hubbard interaction and disorder is included via the one-body hopping operators.  In both the non-interacting and interacting models we make use of recent advances in highly accurate real-space numerical evaluation of topological invariants to compute phase boundaries, and in the non-interacting models determine critical exponents of the transitions. We find different models exhibit distinct stability conditions of the topological phase with respect to interactions and disorder.  We provide a general analytical theory that accurately predicts these trends.



\end{abstract}
\date{\today}
\maketitle

{\it Introduction--}Topological insulators have emerged over the past decade as a topic of intense research focus \cite{Moore:nat10,Hasan:rmp10,Qi:rmp11,Ando:jpsj13}.  When electron-electron interactions are an essential ingredient of a topological phase, the phenomenology becomes remarkably richer, though the full range of possible behaviors is not yet fully understood \cite{Maciejko:np15,Stern:arcmp15,witczak-krempa2014,mesaros2013,Chen:prb13}.   A topic that has thus far received comparatively little attention in the literature is the influence of disorder on interacting topological phases.  For non-interacting topological insulators, there has been a flurry of activity around the so called topological Anderson insulator (TAI),  which results from random terms in the Hamiltonian that drive a non-topological system into a topological state \cite{Girschik:prb15,Xu:prb12,Jiang:prb09,Song:prb12,Jain:prl09,Groth:prl09,Prodan:prb11,Guo:prl10,Prodan:prl10}. However, even in non-topological systems the combination of disorder and interactions is known to be a particularly challenging problem \cite{Imada:rmp98,Lee:rmp06,Georges:rmp96,Basov:rmp11,Dagotto:rmp94}.

In this work we focus on disordered variants of the two-dimensional Kane-Mele (KM) model \cite{Kane:prl05,Kane_2:prl05} that are selected because they do not have a fermion sign problem when simulated with quantum Monte Carlo (QMC) \cite{Hung:prb13,Hung:prb14}, if the disorder is on the hopping terms \cite{Song:prb12,Chua:prb12} (bond disorder) of the model rather than the on-site chemical potential \cite{Jiang:prb09,Jain:prl09,Groth:prl09} (because particle-hole symmetry is preserved in the former case).  These models allow one to investigate the combination of disorder and interactions in a topological system with a high degree of numerical accuracy, especially when combined \cite{Supplemental} with recent advances in the real-space numerical evaluation of topological invariants \cite{ProdanJPhysA2011xk}. 

In the limiting case of weak disorder and interactions, we present an analytical theory that accurately describes the interplay of disorder and interactions near the phase boundary between the topological and trivial phases of the models.  The generality of the analytical theory suggests that it should correctly predict trends in other two and three dimensional models as well, regardless of whether they have a fermion sign problem.  This may open the door to wider studies of disorder in interacting topological phases, at least for certain regimes of the parameter space.  A central result of this work is that the physics of disorder in topological phases depends crucially on the momentum values of gap closing points in the clean limit.  We illustrate this explicitly with our models, which exhibit a different behavior with respect to disorder from those studied previously.

Before turning to the study of interacting models, we first contrast bond and on-site disorder in the non-interacting limit of these models using large scale exact diagonalization and highly accurate real-space evaluation of topological invariants of finite-size systems.  Our computations, combined with a finite-size scaling analysis, allow a precise determination of phase boundaries, as well as the critical exponents of the transition, which agree with general theoretical expectations based on the symmetry classes of the models.


{\it Model Hamiltonians--}In this work we study two variants of the well-known KM model \cite{Kane:prl05,Kane_2:prl05} (as well as the KM model itself),
\begin{eqnarray}
  H_{\textrm{KM}} & = &
 - t\sum_{\langle i,j\rangle,\sigma}c_{i\sigma}^{\dagger}c_{j\sigma}+i \lambda_{SO}\sum_{\langle\langle i,j\rangle\rangle, \sigma} \sigma c^{\dagger}_{i\sigma}\nu_{ij} c_{j\sigma},
\label{eqn:KMHam}
\end{eqnarray}
defined on the honeycomb lattice where $\langle \cdots \rangle$ runs over nearest-neighbor sites and $\langle \langle \cdots \rangle \rangle$ over next-nearest-neighbor sites, and $\sigma=\pm1$ describes the spin projection along the $z$-axis. The operator $c_{i\sigma}$ annihilates a fermion on site $i$, and $c^\dagger_{i\sigma}$ creates a fermion on site $i$. Each unit cell contains two sites, labeled as A and B.  The spin-orbit coupling strength is $\lambda_{SO}$, and $\nu_{ij}=+1$ for counter-clockwise hopping along a hexagonal plaquette and $\nu_{ij}=-1$ for clockwise hopping \cite{Supplemental}.  

 \begin{figure*}[!ht]
{\includegraphics[width=1.6\columnwidth]{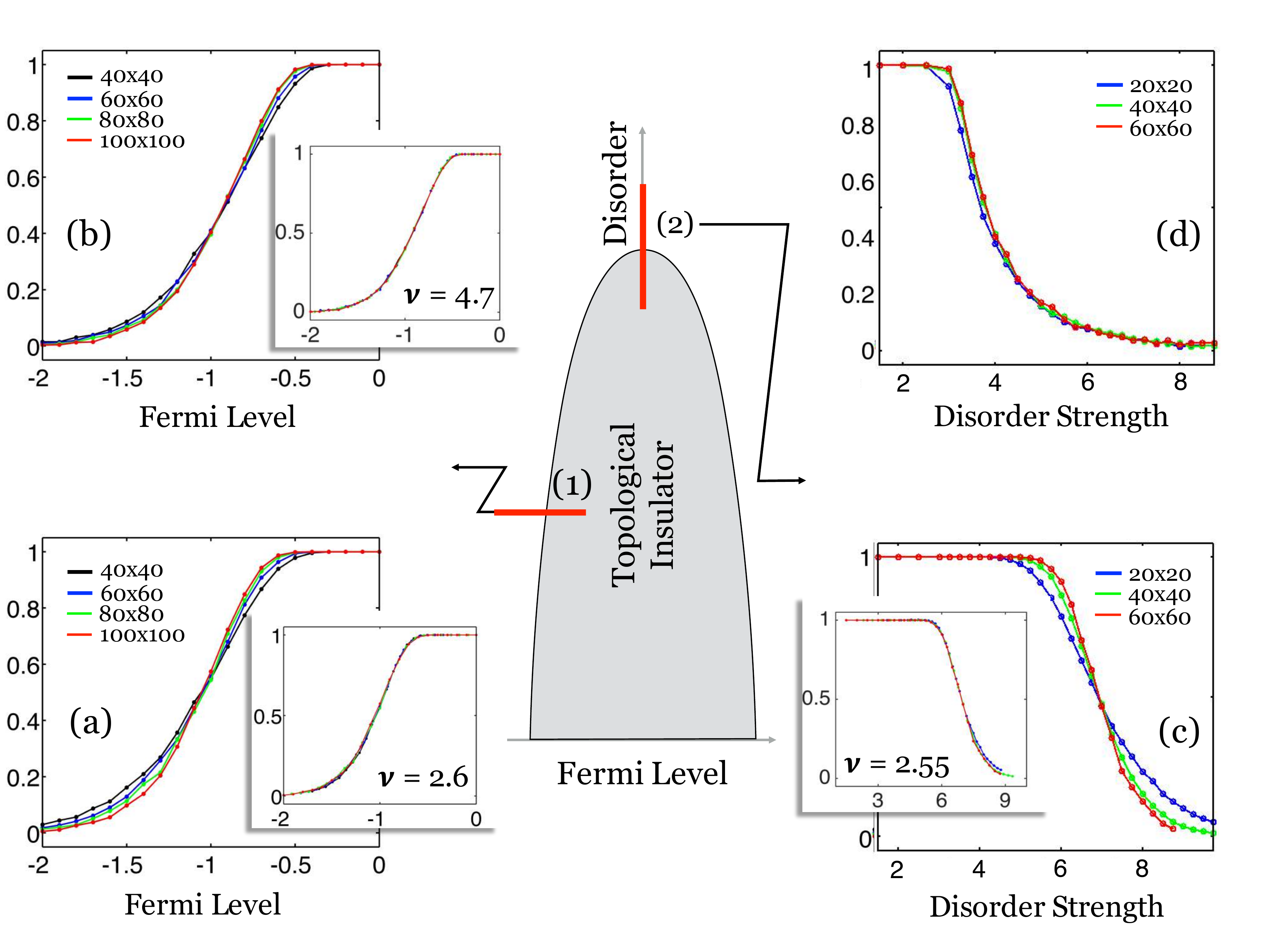}}
\caption{Exact diagonalization study of the Kane-Mele model, Eq.\eqref{eqn:KMHam}, with bond [panels (a) and (c)] and on-site [panels (b) and (d)] disorder. The center of the figure shows a qualitative representation of the phase diagram, with a topological phase ${\rm Ch}=1$ (spin Chern number) trapped inside (gray region) an infinitely thin phase boundary. On the other side of the phase boundary, there is a trivial insulator ${\rm Ch}=0$. For bond-disorder there is an exception at the top of the phase diagram, along the vertical axis, where a metallic phase is stabilized by the particle-hole symmetry. The panels show the evolution of the Chern number along the phase-boundary crossings indicated on the diagram. The various curves in the main panels correspond to different lattice sizes and the insets demonstrate the single-parameter scaling of the data, fit to Eq.\eqref{eq:E_scale} for panels (a) and (b), and Eq.\eqref{eq:W_scale} for panel (c). Case (d) is a transition from an insulating to a metallic state. For further details, see text. }
\label{Fig1}
\end{figure*}

Because the KM model preserves the $z$-projection of the spin, it may be regarded as two ``copies" of the Haldane model \cite{Haldane:prl88}, with
\begin{equation}
  H_{\uparrow}  =  - t\sum_{\langle i,j\rangle}c_{i\uparrow}^{\dagger}c_{j\uparrow}+i \lambda_{SO}\sum_{\langle\langle i,j\rangle\rangle} c^{\dagger}_{i\uparrow}\nu_{ij} c_{j\uparrow}, \label{eqn:H_up}
\end{equation} 
for the spin up component and its time-reversed partner for the spin-down component. For any non-zero value of $\lambda_{SO}$, each spin projection has been shown to be a quantum Hall (QH) insulator \cite{Haldane:prl88} with opposite Chern numbers, ${\rm Ch}=\pm 1$, for the two spin projections.   Likewise, the KM model is in the Z$_2$ topologically non-trivial phase when  $\lambda_{SO}$ is non-zero. Therefore, the Hamiltonian in Eq.\eqref{eqn:KMHam} will be regarded as a ``parent Hamiltonian" to which we will add an on-site Hubbard Interaction, $H_U=\sum_i n_{i \uparrow} n_{i\downarrow}$, other spin-independent hopping terms, and various sources of disorder (both bond disorder and on-site disorder in the non-interacting limit).  Taking advantage of the conservation of the $z$-component of the spin, much of our analysis of the topological properties will be focused on computing the spin Chern number \cite{Sheng:prl06,Prodan:2009oh,Xu_Zhong:prb12}  as a function of model parameters.


{\it Numerical Results: Non-interacting Models--}  Fig.~\ref{Fig1} shows a qualitative phase diagram of the disordered Kane-Mele model, Eq.\eqref{eqn:KMHam}, and the evolution of the topological invariants (spin Chern numbers) as the model transits the phase boundary. We study two types of disorder, on-site disorder $H_{\rm on-site}=\sum_{i} \omega_{i}n_i$ and bond-disorder $H_{\rm bond}=\sum_{\langle i , j \rangle,\sigma} \omega_{ij}c^{\dagger}_{i\sigma}c_{j\sigma}$, which were separately added to the KM model. In both cases, the $\omega$'s are independent random variables drawn uniformly from the interval $[-W/2,W/2]$, with $W$ representing the disorder strength. In Fig.~\ref{Fig1}, $t=1$ and $\lambda_{SO} = 0.3$.

Since the z-component of the spin is conserved, effectively the disordered models belong to the universality class A \cite{Schnyder:prb08}. In this case, the phase boundary consists of an infinitely thin line where the localization length diverges \cite{ProdanJPhysA2011xk}. We point out, however, that at half-filling and with the bond-disorder, the model displays an exact particle-hole symmetry which changes the universality class to D \cite{Schnyder:prb08} and qualitatively modifies the phase diagram. Indeed, in this case it is known that the system will cross into a metallic phase \cite{KagalovskyPRL2008} (similar behavior has been observed for other types of disorder \cite{CastroPRB2015tf}). This strict preservation of the particle-hole symmetry is important for quantum Monte Carlo calculations because the fermion sign-problem can be avoided \cite{HsiangMPL2014,Hung:prb13,Hung:prb14}.

The crucial point of the computations reported in Fig.~\ref{Fig1} is to exemplify a procedure, based on the finite-size critical scaling behavior, for the precise identification of the phase boundary \cite{PriestPRB2014fj}.  To accurately compute the topological invariant, we make use of recent advances in the real-space evaluation of the Chern number, which exhibits exponential convergence when the lattice size exceeds the Anderson localization length \cite{Supplemental}. As the critical boundary is approached, however, the localization length will inherently exceed any conceivable lattice size, hence even with these fast converging algorithms one cannot obtain a sharp stepping profile of the topological invariants. In other words, there will be a sizable region where the value of the topological invariants (of the infinite system) cannot be determined numerically. Nevertheless, we show in the following that the invariants obey finite-size critical scaling behavior which can be used quite effectively to locate the true critical phase boundary and even determine critical exponents.  We averaged over 100 disorder realizations.  

In panels (a) and (b), the phase boundary is traversed as indicated by the red segment (1), hence varying the Fermi level while keeping the disorder strength fixed. Panel (a) corresponds to the on-site disorder of strength $W=4$ while panel (b) to the bond-disorder of strength $W=3$, a lower value because the effect of disorder is stronger in this case (compare panels (c) and (d) to get a sense of how strong). The Chern number was evaluated on finite lattices of increasing sizes and, as expected, the quantization and the transition between the quantized values becomes sharper as the lattice size is increased. The most important feature to notice is that the curves intersect at a single point, signaling the existence of a unique critical point, $E_F^c$, separating two insulating phases. In the insets, we show that, upon the rescaling
\begin{equation}
\label{eq:E_scale}
E_F \rightarrow X = E_F^c + (E_F-E_F^c) \Big ( \frac{L}{L_0} \Big )^\frac{1}{\nu},
\end{equation}   
the curves collapse on top of each other. The optimal collapse is obtained for $E_F^c =-1.00 \pm 0.01$ and $\nu = 2.6 \pm 0.1$ in the case of on-site disorder, and $E_c =-0.98 \pm 0.01$ and $\nu = 4.7 \pm 0.01$ in the case of bond-disorder. Note that the value $\nu = 2.6$ is in very good agreement with the numerical values reported in the literature \cite{SO1,KMO,OSF,FHA,DET,SO2,OGE}.

In panels (c) and (d), the phase boundary is traversed as indicated by the red segment (2), by varying the the disorder strength while keeping the Fermi level fixed at $E_F=0$ (half-filling). In panel (c) we can see again the curves intersecting each other at one single point indicating again the existence of a unique critical point $E_F^c$ separating two insulating phases. In the insets, we show that, upon the rescaling
\begin{equation}
\label{eq:W_scale}
W \rightarrow X = W^c + (W-W^c) \Big ( \frac{L}{L_0} \Big )^\frac{1}{\nu},
\end{equation}   
the curves collapse on top of each other. The optimal collapse is obtained for $W^c = 6.75 \pm 0.01$ and $\nu = 2.55$, with the small differences in the critical exponent from case (a) attributed to the smaller lattices in the simulations. 

Panel (d), which corresponds to the bond-disorder, looks markedly different from the previous panels. Here, the curves decay to zero without intersecting each other at a unique critical point. In fact, on the larger $W$ side, there is little variation of the curves from one system-size to another, which is an indicative that the system entered a metallic phase. The scaling analysis should apply over the insulating side of the phase diagram, but it was difficult to perform in this case and we omitted it in Fig.~1. Nevertheless, the analysis indicates that the critical point, which marks the phase boundary between the topological insulator and metallic phase, is located very  close to the point where the graphs start to deviate from the quantized value of 1.

{\it Numerical Results: Interacting Models--}We study two generalizations of the Kane-Mele-Hubbard (KMH) model, $H_{\textrm{KMH}}=H_{\textrm{KM}}+H_U$ \cite{Wu:prb15,Laubach:prb14,Reuther:prb12,Griset:prb12,Zheng:prb11,Yu:prl11,Rachel:prb10,Hohenadler:jpc2013,Hohenadler:prb12,Parisen:prb15},  known as the generalized KMH (GKMH) model and the dimerized  KMH (DKMH) model. In the absence of disorder, both models are already well characterized by QMC \cite{Hung:prb13,Hung:prb14}, dynamical mean-field theory \cite{Chen_Yao:prb15}, and analytical methods \cite{Lai:ijmpb15,Lai:prb14}, therefore there is a solid starting point for investigating the effects of disorder on interacting topological systems. 

 A critical difference between the KM model and the non-interacting GKM and DKM models is that that gap closings occur at the ${\bf M}$ points of the Brillouin zone in the latter two, while it occurs at the ${\bf K}$ and ${\bf K'}$ points in the former \cite{Supplemental}.  This has important consequences for how the bond and the on-site disorders influence the phase transitions in the GKM and DKM models \cite{Supplemental}.  In particular, for models with a gap closing at the ${\bf K}$ and ${\bf K'}$ points, bond disorder and on-site disorder lead to {\rm different} signs of contributions to the gap \cite{Song:prb12}.  By contrast, when the gap closings are at the ${\bf M}$ points, bond disorder and on-site disorder lead to the {\rm same} sign of contribution to the gap \cite{Supplemental}.  Therefore, one expects qualitatively similar results for the disordered GKMH and DKMH models regardless of whether one studies bond disorder or on-site disorder, so we make use of the fact that we can simulate the former in QMC without a sign problem.

 The GKMH model is given by $H_{\textrm{GKMH}}=H_{\textrm{KMH}}+H_{t_3}$, where $H_{t_3}=- t_{3}\sum_{\langle \langle \langle i,j\rangle \rangle \rangle,\sigma} c_{i\sigma}^{\dagger}c_{j\sigma}$ where $\langle \langle \langle \cdots \rangle \rangle \rangle$  runs over third-nearest-neighbor sites, and the DKMH model is given by $H_{\textrm{DKMH}}=H_{\textrm{KMH}}+H_{t_d}$ where $H_{t_d}=\delta t_{d}\sum_{i\in A,j=i+\hat{e}_3,\sigma} c_{i\sigma}^{\dagger}c_{j\sigma}$ where we chose the dimerized bonds along the nearest neighbor $\hat{e}_3$ direction \cite{Supplemental}.  In the non-interacting limit of the GKMH model, $t_3$ drives a quantum phase transition at $t_3=1/3$, with a $Z_2$ topological insulator with spin Chern number $\pm 1$ for $0<t_3<1/3$ and a trivial insulator for $t_3>1/3$ with spin Chern number $\mp 2$ \cite{Hung:prb13,Hung:prb14}. In the non-interacting limit of the DKMH model, $\delta t_d$ drives a topological phase transition at $\delta t_d=1$, with spin Chern number $\pm 1$ for $0<\delta t_d<1$ and a trivial insulator for $\delta t_d>1$ with spin Chern number 0 \cite{Hung:prb13,Hung:prb14}.  The one-body parameters $t_3$ and $\delta t_d$ also drive a quantum phase transition at finite Hubbard $U$ (for $U$ below the critical value for a magnetic transition, as we consider in this work \cite{Hung:prb13,Hung:prb14,Chen_Yao:prb15,Lai:ijmpb15,Lai:prb14}), though the critical values of  $t_3$ and $\delta t_d$ generally depend on the strength of the interaction \cite{Hung:prb13,Hung:prb14,Chen_Yao:prb15,Lai:prb14,Lai:ijmpb15}.

\begin{figure}[ht]
\epsfig{file=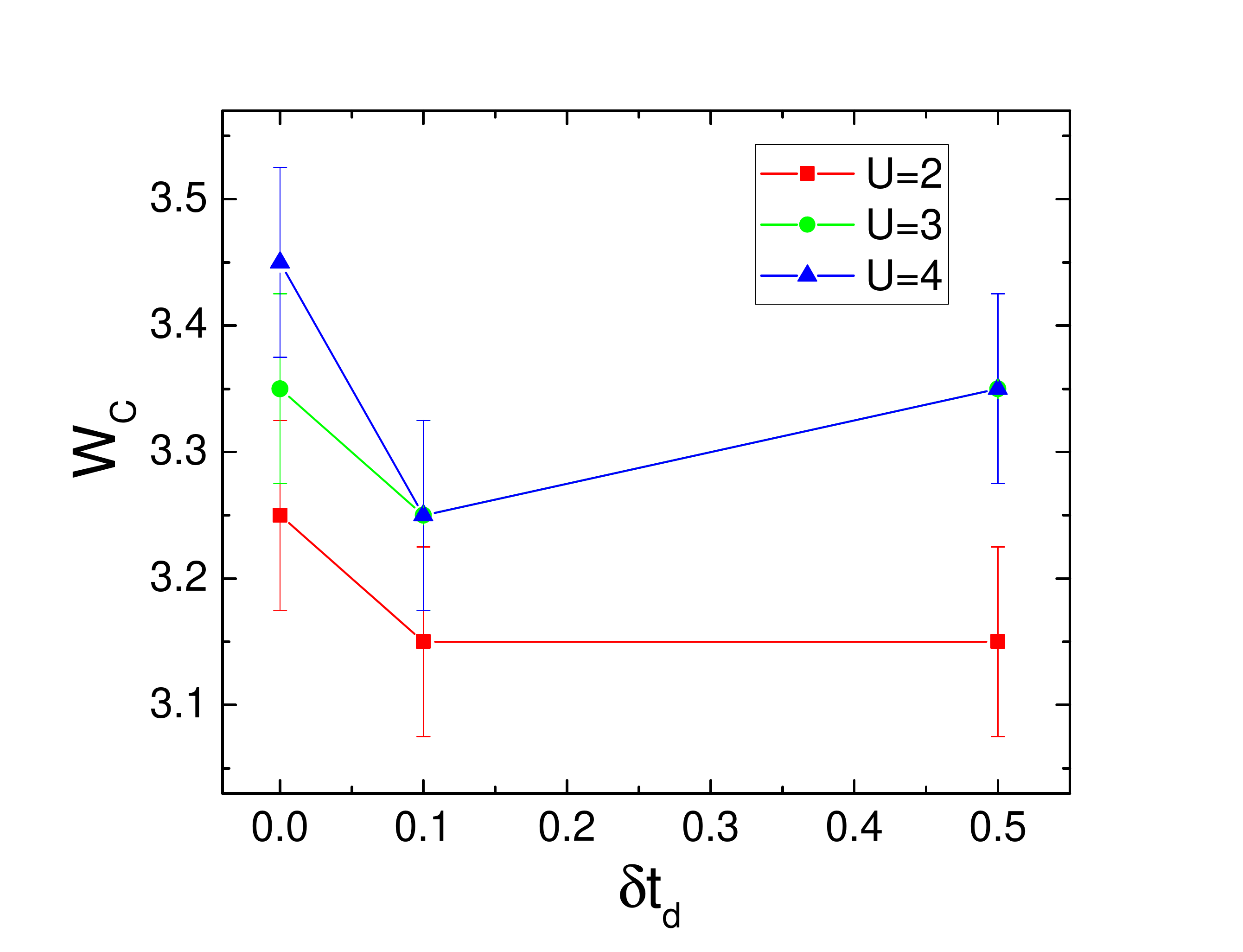,clip=0.1,width=0.49\linewidth,angle=0}
\epsfig{file=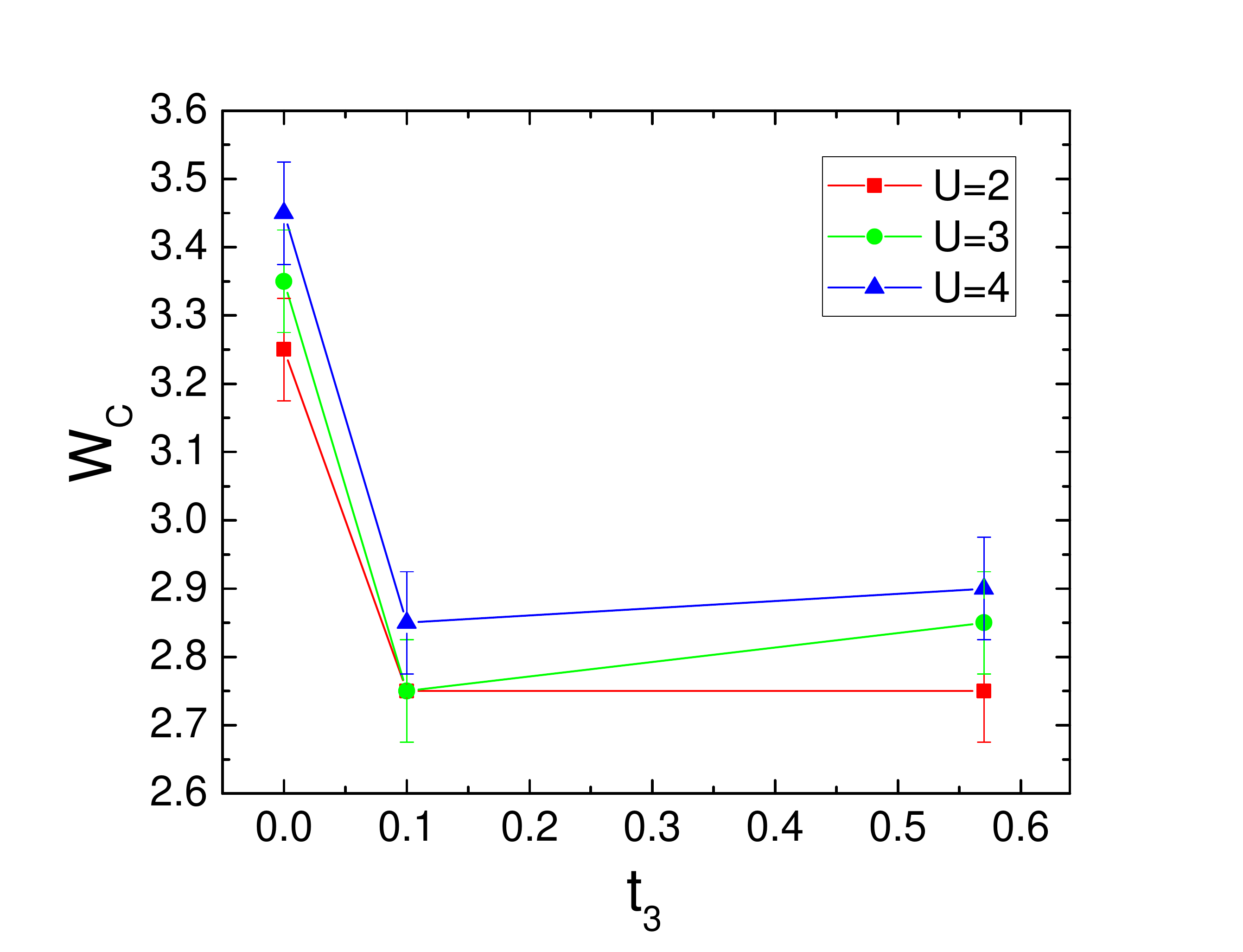,clip=0.1,width=0.49\linewidth,angle=0}
\epsfig{file=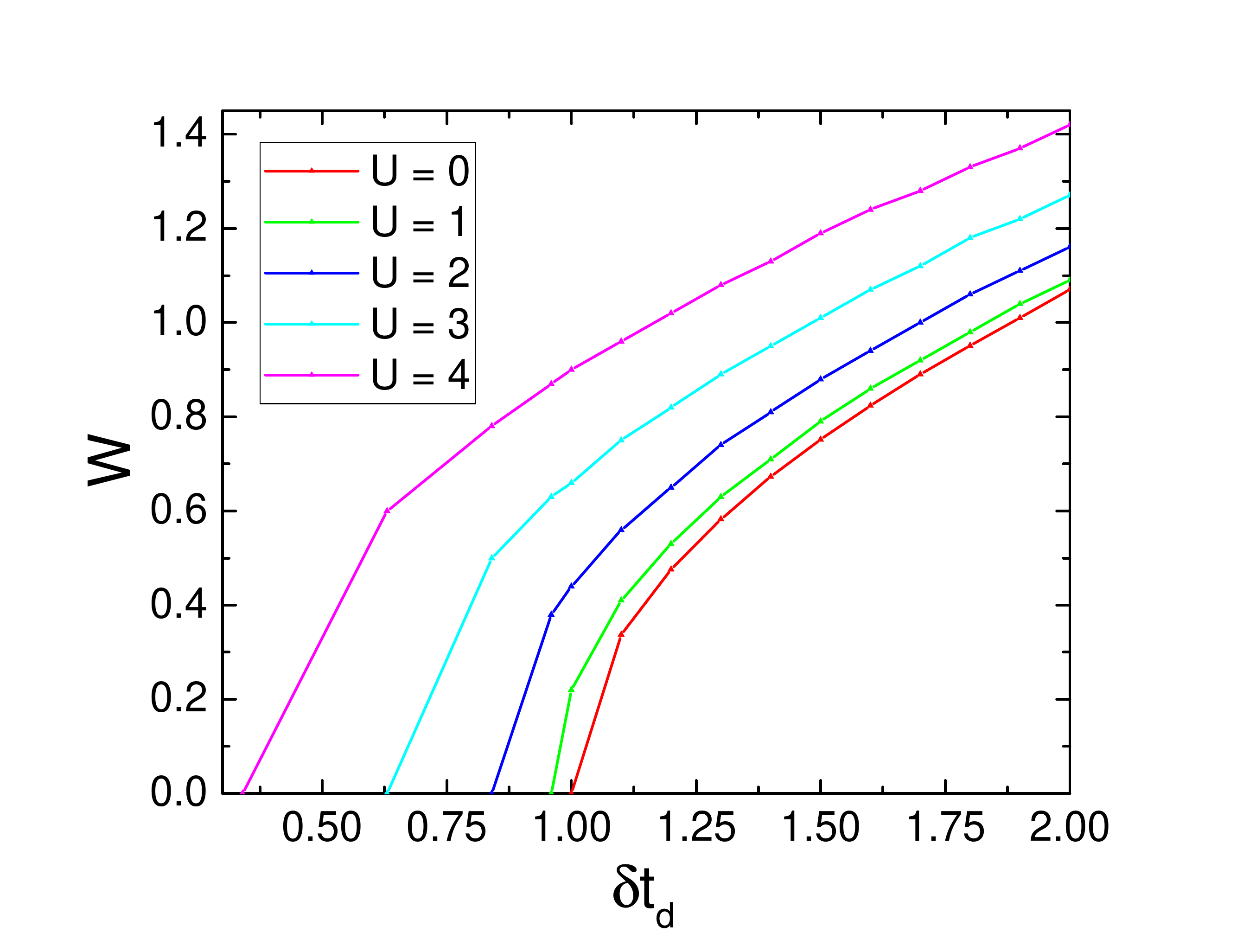,clip=0.1,width=0.49\linewidth,angle=0}
\epsfig{file=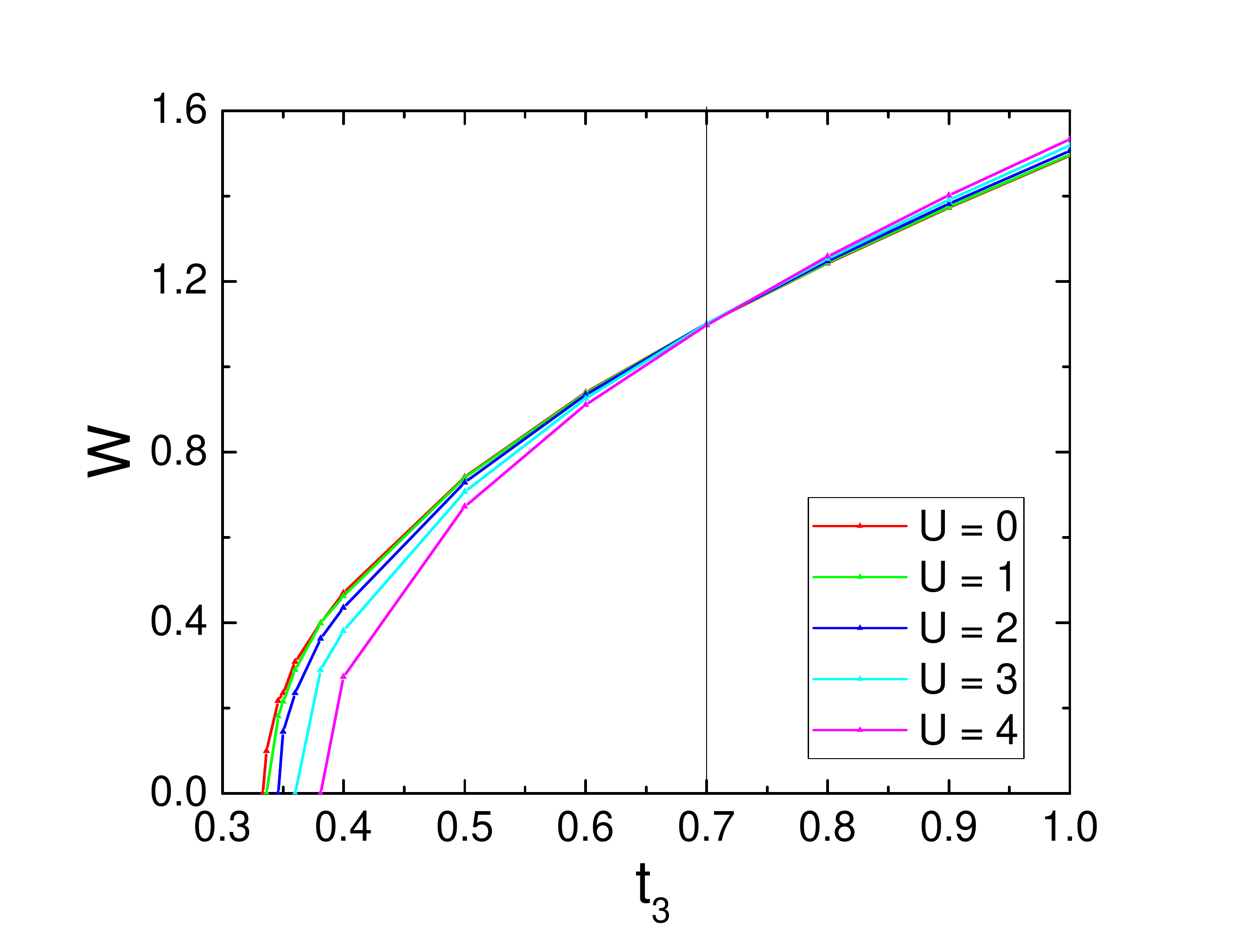,clip=0.1,width=0.49\linewidth,angle=0}
\caption{{\bf Top Panels:} Influence of disorder and interactions on the phase diagram of the DKM and GKM models computed within QMC.  The critical disorder strength is determined by analyzing the disorder averaged spin Chern number \cite{Supplemental}. {\bf Bottom Panels:}  Influence of disorder and interactions on the phase diagram of the DKMH and GKMH models computed using the self-consistent Born approximation (SCBA) \cite{Supplemental}. The curves represent the critical phase boundaries for various values of $U$ separating the $Z_2$ topological/trivial phases, located in the upper/lower regions, respectively.  Disorder tends to stabilize the topological phases in both models. Interactions tend to destabilize the topological phase in the DKMH model and stabilize it in the GKMH model.} 
\label{fig:disorder_boundaries_QMC}
\end{figure}

We also analyze the disordered GKMH and DKMH models system using the self-consistent Born approximation (SCBA) \cite{Song:prb12,Jiang:prb09,Jain:prl09,Groth:prl09}.  We are interested in exploring the interplay of topology, interactions, and disorder on an equal footing.  The QMC calculations are most stable for moderate interaction strengths, $U$, but we are able to analytically study the weak interaction limits using perturbation theory (in the interactions) combined with mean-field theory to obtain a renormalized band structure \cite{Lai:ijmpb15,Lai:prb14} that is used in the SCBA.   Prior work has shown that (i) in the weak disorder limit, the SCBA provides a high level of quantitative agreement with exact diagonalization studies \cite{Groth:prl09}, and (ii) our pertubation+mean-field theory works semi-quantitatively well when compared to QMC \cite{Lai:ijmpb15,Lai:prb14}.  We therefore combine the two to provide an analytical theory of combined disorder and interaction effects in topological insulators \cite{Supplemental}, generally applicable to any model, regardless of whether it can be simulated with fermion sign-free QMC.

Our main QMC and SCBA results are shown in Fig.~\ref{fig:disorder_boundaries_QMC}. In the top panels, the critical disorder strength is shown to drive the spin Chern number for each value of $U,t_3/\delta t_d$ to zero \cite{Supplemental}.  These computationally demanding results make use of highly accurate real-space Chern number evaluation \cite{Supplemental}, but can only be performed for one ``large" lattice size, preventing a finite-size scaling analysis and smaller error bars.  Overall, the trend is that larger interactions tend to lead to a larger $W_c$, which is consistent with the SCBA in the lower panels, for much of the parameter space.  An important result to emerge from the SCBA in the lower panels is that disorder tends to stabilize the TI state which lives ``above" the curves in the lower panel, consistent with analytical arguments \cite{Supplemental}.


{\it Conclusions--}In this work we have studied the interplay of disorder and interactions in two-dimensional lattice models that exhibit a $Z_2$ topological insulator state.  In addition, we have also used recent advances in the real-space numerical evaluation of topological invariants to accurately compute the scaling behavior across the phase boundaries of the disordered Kane-Mele model.  The values of the critical exponents agree with theoretical expectations.  Bond and on-site disorder were contrasted, which exhibited different scaling exponents in the Kane-Mele model, see Fig.~\ref{Fig1}.

For disordered interacting models, we focused on two generalizations of the Kane-Mele-Hubbard model--both of which preserve particle-hole symmetry at half-filling, even when bond disorder is included.  This allowed for a fermion sign problem-free QMC study that could be compared with a perturbative (in disorder and interaction strength) self-consistent Born approximation calculation (detailed in \cite{Supplemental}), see Fig.~\ref{fig:disorder_boundaries_QMC}.  To within the numerical uncertainties of the QMC, there is agreement in the trends.  Moreover, the extended (for interactions) SCBA we applied can be used to study the interplay of disorder and interactions in a completely general context, independent of whether the lattice modal has a fermion sign problem.  

An important feature of the generalized Kane-Mele models we study is that the gap closing is not at the ${\bf K}$ and ${\bf K'}$ points (or the ${\bf \Gamma}$ point) of the Brillouin zone, but rather the ${\bf M}$ points, which essentially presents a new class of models where disorder can be studied along with interactions.  The gap closings at the ${\bf M}$ points implies that on-site and bond disorder will have the same effect on the sign of the gaps near the gap closing points, in contrast to other models in the literature.  We hope this work will stimulate further theoretical and experimental study  on disordered, interacting, topological systems. 

\noindent {\it Acknowledgements.} E.P. acknowledges support from U.S. NSF grant DMR-1056168. G.A.F. acknowledges support from ARO grant W911NF-14-1-0579 and NSF DMR-1507621.

%

\newpage

\
\newpage

\includepdf[pages=1]{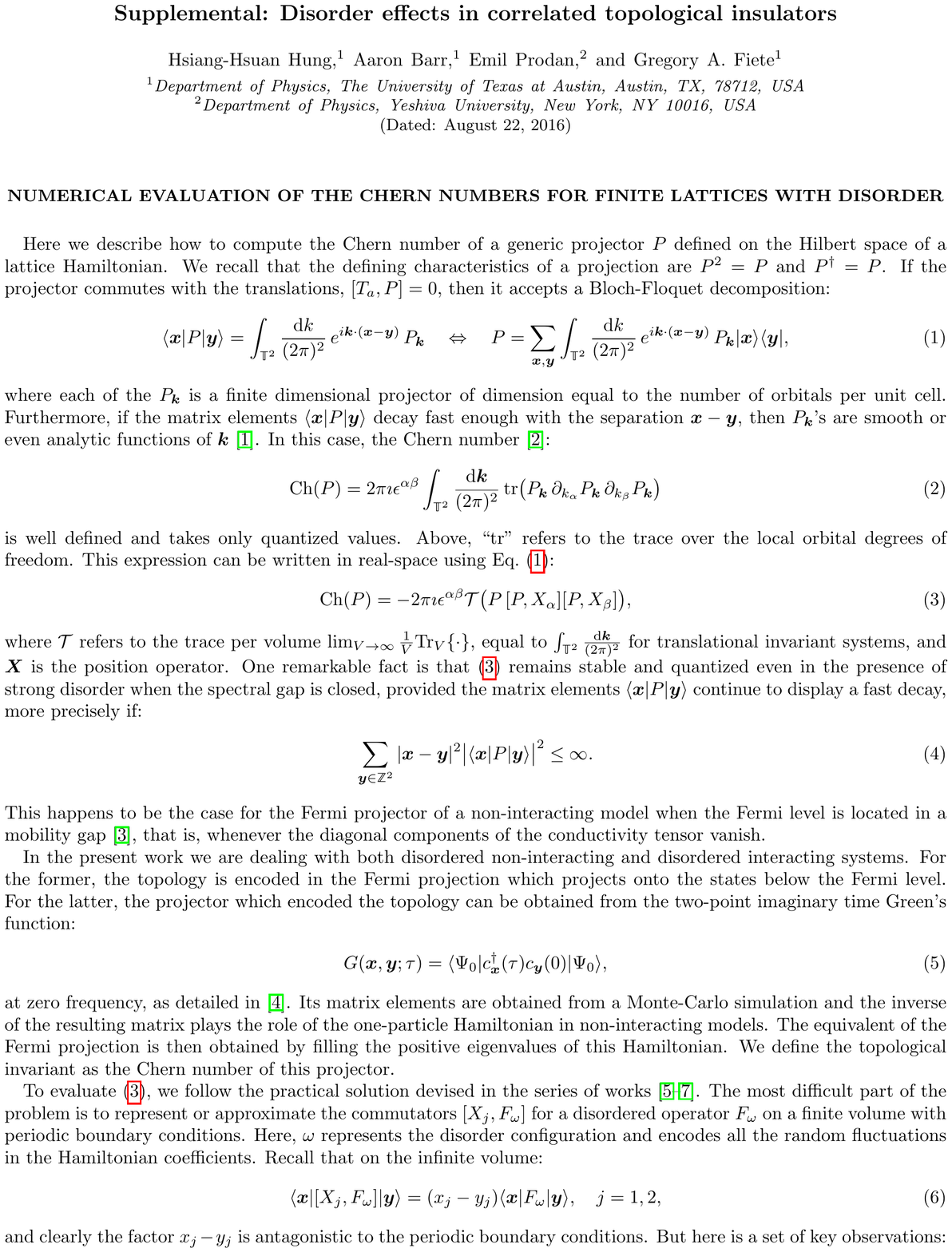}
 \
\newpage 

\includepdf[pages=2]{Supplemental-8-22-16.pdf}
\
\newpage

\includepdf[pages=3]{Supplemental-8-22-16.pdf}
\
\newpage

\includepdf[pages=4]{Supplemental-8-22-16.pdf}
\
\newpage

\includepdf[pages=5]{Supplemental-8-22-16.pdf}
\
\newpage

\includepdf[pages=6]{Supplemental-8-22-16.pdf}
\
\newpage

\includepdf[pages=7]{Supplemental-8-22-16.pdf}
\
\newpage

\includepdf[pages=8]{Supplemental-8-22-16.pdf}

\end{document}